\documentclass[journal=jacsat,manuscript=article]{achemso}
\usepackage[version=3]{mhchem}
\usepackage{tabularx}
\usepackage{float}
\usepackage{amssymb}
\usepackage{color}
\usepackage[dvipsnames]{xcolor}
\usepackage{amsmath}

\author{Daphna Shimon}
\affiliation{Institute of Chemistry, The Hebrew University of Jerusalem, Edmond J. Safra, Givat Ram, Jerusalem, Israel.}
\email{daphna.shimon@mail.huji.ac.il}
\author{Kelly A. Cantwell}
\affiliation{Department of Physics and Astronomy, Dartmouth College, Hanover, NH 03755, U.S.A.}
\author{Linta Joseph}
\affiliation{Department of Physics and Astronomy, Dartmouth College, Hanover, NH 03755, U.S.A.}
\author{Chandrasekhar Ramanathan}
\affiliation{Department of Physics and Astronomy, Dartmouth College, Hanover, NH 03755, U.S.A.}

\title{Room Temperature DNP of Diamond Powder Using Frequency Modulation}

\begin{document}

\begin{abstract}

\noindent{}Dynamic nuclear polarization (DNP) is a method of enhancing NMR signals via the transfer of polarization from electron spins to nuclear spins using on-resonance microwave (MW) irradiation. In most cases, monochromatic continuous-wave (MCW) MW irradiation is used. Recently, several groups have shown that the use of frequency modulation of the MW irradiation can result in an additional increase in DNP enhancement above that obtained with MCW. The effect of frequency modulation on the solid effect (SE) and the cross effect (CE) has previously been studied using the stable organic radical 4-hydroxy TEMPO (TEMPOL) at temperatures under 20 K. Here, in addition to the SE and CE, we discuss the effect of frequency modulation on the Overhauser effect (OE) and the truncated CE (tCE) in the room-temperature $^{13}$C-DNP of diamond powders. We recently showed that diamond powders can exhibit multiple DNP mechanisms simultaneously due to the heterogeneity of P1 (substitutional nitrogen) environments within diamond crystallites. We explore the enhancement obtained via the two most important parameters for frequency modulation  1) Modulation frequency, $f_{m}$ (how fast the modulation frequency is varied) and 2) Modulation amplitude, $\Delta\omega$ (the magnitude of the change in microwave frequency) influence the enhancement obtained via each mechanism. Frequency modulation during DNP not only allows us to improve DNP enhancement, but also gives us a way to control which DNP mechanism is most active. By choosing the appropriate modulation parameters, we can selectively enhance some mechanisms while simultaneously suppressing others. 

\end{abstract}

\section{Introduction}

Dynamic nuclear polarization (DNP) enhances NMR signals by transferring spin polarization from electrons to nuclei by on-resonance microwave (MW) irradiation of the electron spins \cite{c1,c2,c3}. In most cases, monochromatic continuous-wave (MCW) MW irradiation is used and the single frequency of the MW irradiation is chosen to maximize the DNP enhancement \cite{c1,c2,c3}.  A DNP spectrum (the DNP enhancement measured as a function of MW frequency) is typically recorded to determine the optimal MW frequency for the MCW. The shape of the DNP spectrum depends on the nature of the electron spin, the nucleus that is being enhanced, and the DNP mechanism involved \cite{c4,c5,c6,c7}.

Recently, several groups have demonstrated that employing MW irradiation schemes that are more complex than simple MCW irradiation can result in an additional increase in the DNP enhancement (above that obtained with MCW). \cite{c15,c16,c17,c18,c19,Hovav2014,c21,c22,c23,c24,c25,c26,c27} One method that works well at high magnetic fields ($>$ 3.3 T) and for the case of relatively limited MW power is frequency modulation, or trains of broadband chirp pulses. Under static (non-spinning) conditions, this irradiation can improve the DNP performance several fold compared to MCW irradiation \cite{c15,c16,c17,c18,c19,Hovav2014,c21,c22,c23,c24,c25,c26,c27}. The increase in the enhancement is usually attributed to the fact that a larger number of electrons can be excited with broadband MW irradiation  \cite{Hovav2014,c26}. However, the size of the additional  enhancement depends on many experimental factors, such as the modulation frequency $f_m$ (i.e. the rate of change of the MW frequency), the modulation amplitude $\Delta\omega$ (the bandwidth of the frequency change), the nuclei involved, the radical type and other physical parameters \cite{Hovav2014,c26,c27}. Several groups have shown that for a given sample, the improvement of the DNP enhancement is maximal (optimal) at certain combinations of $\Delta\omega$ and $f_m$, with other values often resulting in sub-optimal DNP enhancements \cite{Hovav2014,c27}.

The effect of frequency modulation on the solid effect (SE) and the cross effect (CE) has previously been studied using the stable organic radical TEMPOL at temperatures under 20 K \cite{Shimon2022}. We showed that the maximum of the DNP spectrum for the SE and the CE depends both on the modulation amplitude and on electron spectral diffusion (eSD), which spreads the electron depolarization throughout the EPR line. We also showed that modulation results in a transition from SE to CE enhancement in those model samples.\cite{Shimon2022} To come to this conclusion, it was necessary to update the original DNP model for large spin systems (known as the indirect CE model) by Hovav et al.\ to include the effects of frequency modulation \cite{Hovav2014}. 

Here, in addition to the SE and CE, we discuss the effect of frequency modulation on the Overhauser effect (OE) and the  truncated CE (tCE) in room-temperature $^{13}$C-DNP experiments on diamond powders.   
We recently showed that diamond powders can exhibit multiple DNP mechanisms simultaneously due to the heterogeneity of P1 environments within diamond crystallites, while also enabling a greater than 100-fold enhancement of the $^{13}$C signal \cite{diamond_1_2022_arxiv}. 

The P1 center is a substitutional nitrogen defect (spin-1/2) in diamond that has been used to hyperpolarize nuclei at both low and high magnetic fields. Previously, P1 centers have been shown to produce a significant increase in the nuclear spin polarization during DNP at cryogenic temperatures \cite{Lock1992, Reynhardt1998,Reynhardt2001,Casabianca2011,Bretschneider2016,Kwiatkowski2018}. Bretschneider et al.\ also reported a room temperature $^{13}$C DNP enhancement of 130 at 9.4 T under 8 kHz MAS conditions with 10 W of microwave power at 263 GHz \cite{Bretschneider2016}.  

Since all four mechanisms are observed in the same diamond powder sample, we are able to examine and compare how the modulation frequency and  modulation amplitude influence the DNP enhancement obtained via each of these mechanisms.
Frequency modulation during DNP allows us to improve DNP enhancement. More importantly, it gives us a way to control which DNP mechanism is most active, by selectively enhancing some mechanisms and suppressing other mechanisms, depending on the modulation parameters.

\section{Results and Discussion}

We performed frequency modulated DNP experiments on a diamond powder sample at 3.34 T.  Details about the sample and experimental methods can be found in the Methods Section at the end of the paper. This powder sample exhibits four DNP mechanisms, the SE, CE, OE and tCE, which we identified in a previous publication by fitting a non-modulated DNP spectrum \cite{diamond_1_2022_arxiv}.

\begin{figure}[htb]
    \centering
    \includegraphics[width=0.49\textwidth]{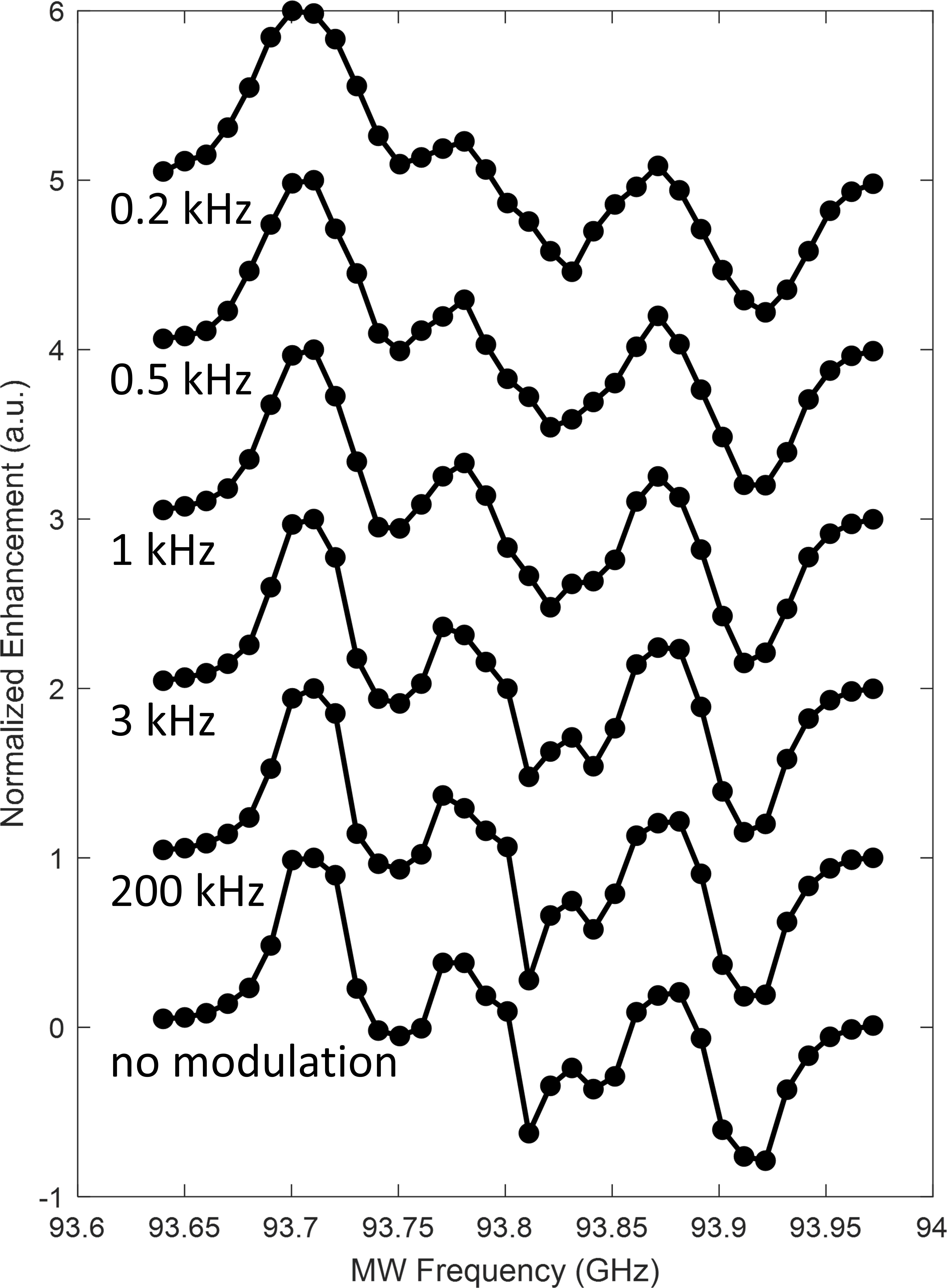}
    \caption{$^{13}$C-DNP spectra, showing normalized DNP enhancement as a function of the MW irradiation frequency at different modulation frequencies, $f_{m}$, as indicated in the figure. The same modulation amplitude, $\Delta\omega=50$ MHz, and build-up time (120 s) were used in all experiments.}
    \label{fig:DNPsweep_vs_f}
\end{figure}

In Figure \ref{fig:DNPsweep_vs_f}, we compare the DNP spectrum acquired with a constant modulation amplitude ($\Delta\omega=50$ MHz), at various modulation frequencies, $f_{m}$, ranging from 200 Hz to 200 kHz. Each spectrum is acquired by changing the central irradiation frequency while applying the same modulation parameters and measuring the DNP enhancement across the central frequency range of interest for the diamond powder sample. The figure shows that the shape of the DNP spectrum measured at large modulation frequencies looks identical to that without modulation. As the modulation frequency is lowered, we see that the sharper features of the DNP spectrum begin to broaden and blur, and that this broadening increases as the $f_m$ is decreased. The enhancement increased by around 20\% at $f_{m}=$3 kHz and then decreased back to its original value as $f_{m}$ is decreased further.

\begin{figure}[htb]
    \centering
    \includegraphics[width=1\textwidth]{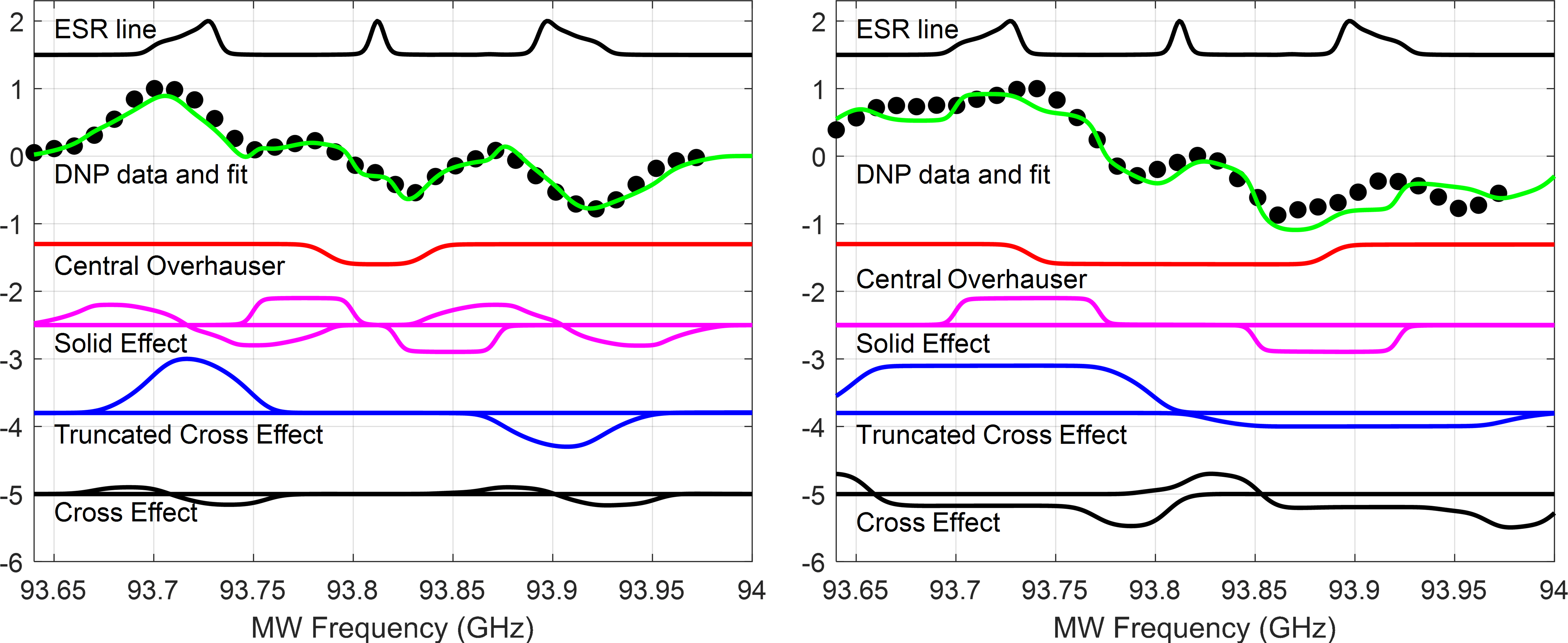}
    \caption{Fit of experimental constant-frequency DNP spectrum (black circles) of the diamond powder sample using a sum (green line) of the OE (red line), SE (magenta line), the truncated CE (blue line), and the CE (black line) for (left) $\Delta\omega=50$ MHz and (right) $\Delta\omega=50$ MHz. The DNP build-up time was 120 s.}
    \label{fig:DNPsweepfits}
\end{figure}

Next, we kept the modulation frequency constant (at 200 Hz) and observed how the modulation amplitude affected the DNP spectrum. In Figure \ref{fig:DNPsweepfits} we compare the DNP spectra acquired with $\Delta\omega=50$ MHz (also shown in Figure \ref{fig:DNPsweep_vs_f}) and $\Delta\omega=150$ MHz. We also fit these DNP spectra using the same model we used in our previous publication on diamond DNP \cite{diamond_1_2022_arxiv}, which is described in the Methods section. Increasing $\Delta\omega$ results in a large change to the DNP spectrum.  Fitting the data allows us to identify two key effects: (i) blurring of the features due to the modulation, and (ii) changes in the relative intensities of each DNP mechanism due to the modulation. We see that with a larger $\Delta\omega$, the contribution of the SE decreases, while the contribution of the CE and tCE increases, which match our previous results \cite{Shimon2022}.

In order to better understand these observations we must try and separate the various DNP mechanisms. This is not an easy task, as the mechanisms overlap in many cases. However, we have identified specific frequencies that mark the maximum DNP enhancement from each mechanism, and contain the least `contamination' from other mechanisms. These frequencies were chosen based on the fit of the non-modulated DNP spectrum \cite{diamond_1_2022_arxiv}. The frequencies are: $\omega$(OE) = 93.814 GHz, $\omega$(SE) = 93.777 GHz, $\omega$(tCE) = 93.725 GHz, and $\omega$(CE) = 93.702 GHz. Note, however, that the overlap between mechanisms grows larger as the frequency modulation amplitude is increased. 

When thinking about how frequency modulation affects different DNP mechanisms it is helpful to consider two different cases: 1) small spin systems, where there are only one or two electron spin packets; and 2) large spin systems, where there are many electron spin packets. Small spin systems featuring the SE and the CE under modulation have previously been described,\cite{Hovav2014} and here we expand the theory to include the OE and tCE in small spin systems.
In large spin systems, the addition of frequency modulation will increase the number of electron spins participating in the DNP process, because, instead of irradiating a single spin-packet, we are able to irradiate many more electron spins packets \cite{Hovav2014, Shimon2021, Shimon2022}. By irradiating more electrons, we see an increase in the DNP enhancement, an important reason why frequency modulation is employed \cite{Hovav2014, Shimon2021, Shimon2022}. This is, in principle, true for all DNP mechanisms. However, when considering large spin systems, we must also consider the effect of electron-electron spectral diffusion (eSD) on the DNP process. eSD can spread electron depolarization throughout the EPR line, and thus also affects the DNP enhancement, by affecting the amount of electron polarization available for DNP \cite{Shimon2022}.  Additionally, because each DNP mechanism behaves differently under frequency modulation, each mechanism must be considered separately, and there is often a balance between involving more electrons to increase the enhancement and saturating additional electron or electron-nuclear transitions that decrease the enhancement. 

\subsection{Solid Effect and Cross Effect}

\subsubsection{Experimental Results}
\begin{figure}[H]
    \centering
    \includegraphics[width=0.99\textwidth]{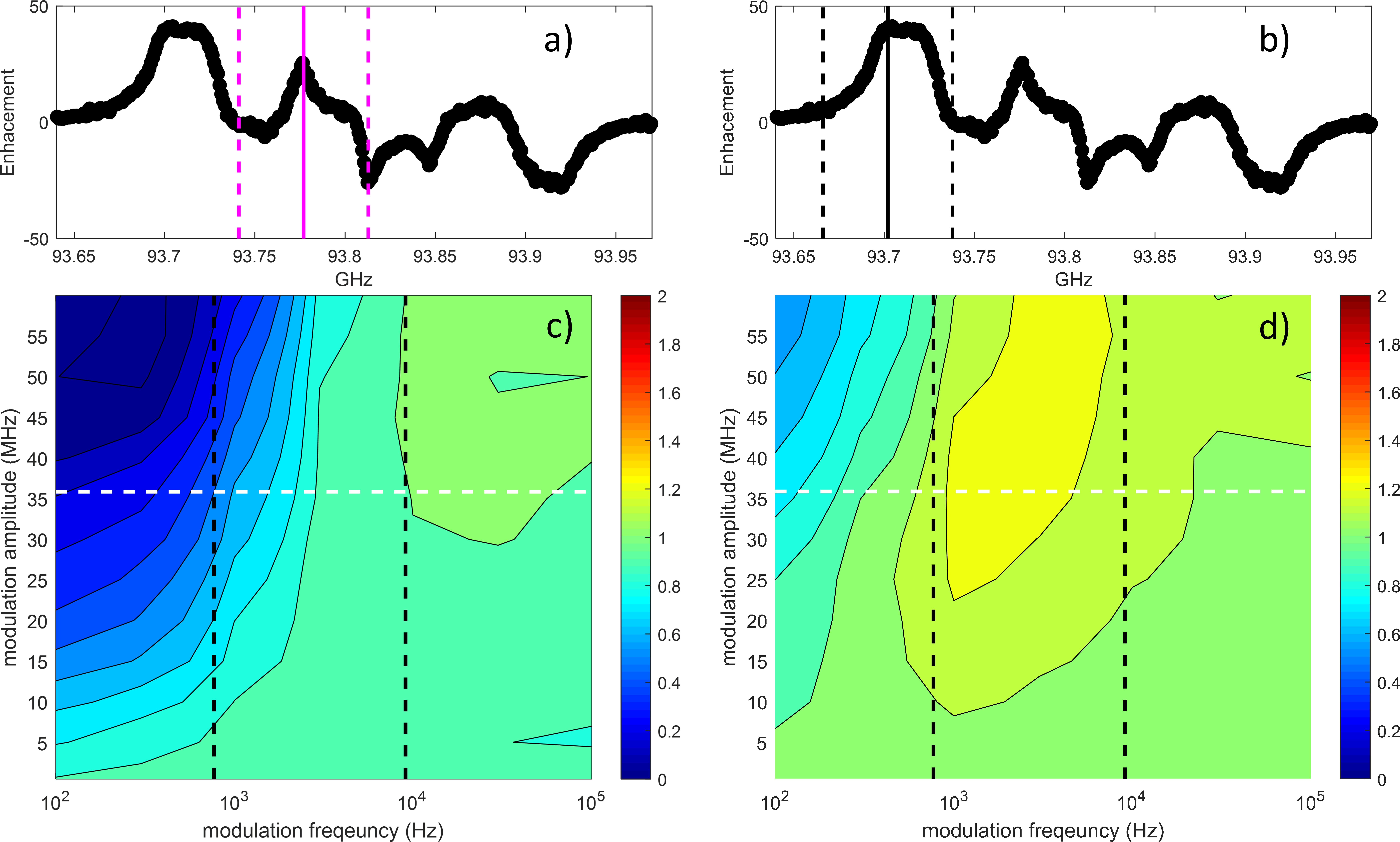}
    \caption{Non-modulated experimental DNP spectra showing the position of the central irradiation frequency (black symbols), $\omega_\text{MW}$, for the a) SE (solid black line) and b) CE (solid magenta line) mechanisms. The dashed lines represent the edges of the MW irradiation when $\Delta\omega=\omega_{C}$.
    DNP enhancement at the c) SE and d) CE frequencies as a function of the modulation frequency, $f_{m}$, and the modulation amplitude, $\Delta\omega$. The dashed white line marks the position where $\Delta\omega=\omega_{C}$. The black dashed lines represent the two timescales of T$_{1e}$ values of a bi-exponential fit of an inversion-recovery experiment using pulsed EPR measured at 2.5 GHz, previously published.\cite{diamond_1_2022_arxiv} }
    \label{fig:contours_SE_CE}
\end{figure}

We begin by exploring the effect of changing the modulation amplitude and frequency on the DNP enhancement at the SE and the CE frequencies.  Figures \ref{fig:contours_SE_CE}(a) and (b) show the non-modulated DNP spectrum and the locations of the center frequencies (solid blue line) of the MW irradiation ($\omega_\text{MW}$) used for the SE and CE respectively. The blue dashed lines in these figures mark the frequency range that corresponds to a modulation amplitude of $\Delta\omega=\omega_C$, the $^{13}$C Larmor frequency, on either side of $\omega_\text{MW}$.  

Figure \ref{fig:contours_SE_CE}(c) and (d) show the effect of changing the modulation amplitude and frequency at the SE and CE frequencies respectively. The colors represent the ratio of the DNP enhancement with frequency modulation to that measured without frequency modulation. The black dashed lines in panels (c) and (d) correspond to the two T$_{1e}$ values (110 $\mu$s and 1.3 ms) measured at room temperature at 2.5 GHz \cite{diamond_1_2022_arxiv}. The shorter timescale (right line) likely represents reorganization of the electron polarization as a result of eSD, or the recovery time of clusters of P1 centers or other paramagnetic defects. The long timescale (left line) is likely to be the recovery time of isolated P1 centers. We did not adjust these values despite the fact that the DNP was performed at a higher field. The dashed white line marks position where the modulation amplitude is equal to $\omega_{C}$.

The key features that we observe are: (i)  The SE is very sensitive to the combination of low modulation frequencies f$_m$ and large modulation amplitudes $\Delta\omega$; (ii) Frequency modulation results only in a very small improvement of the SE enhancement (from $\sim 26$ to $\sim 27$), and in most cases results in a decrease in the enhancement; and (iii) The CE shows a range of optimal conditions where there is an increase of about $\sim 20 \%$ to the enhancement (from $\sim 40$ to $\sim 51$).

\subsubsection{Small spin system simulations}
\begin{figure}[H]
    \centering
    \includegraphics[width=0.99\textwidth]{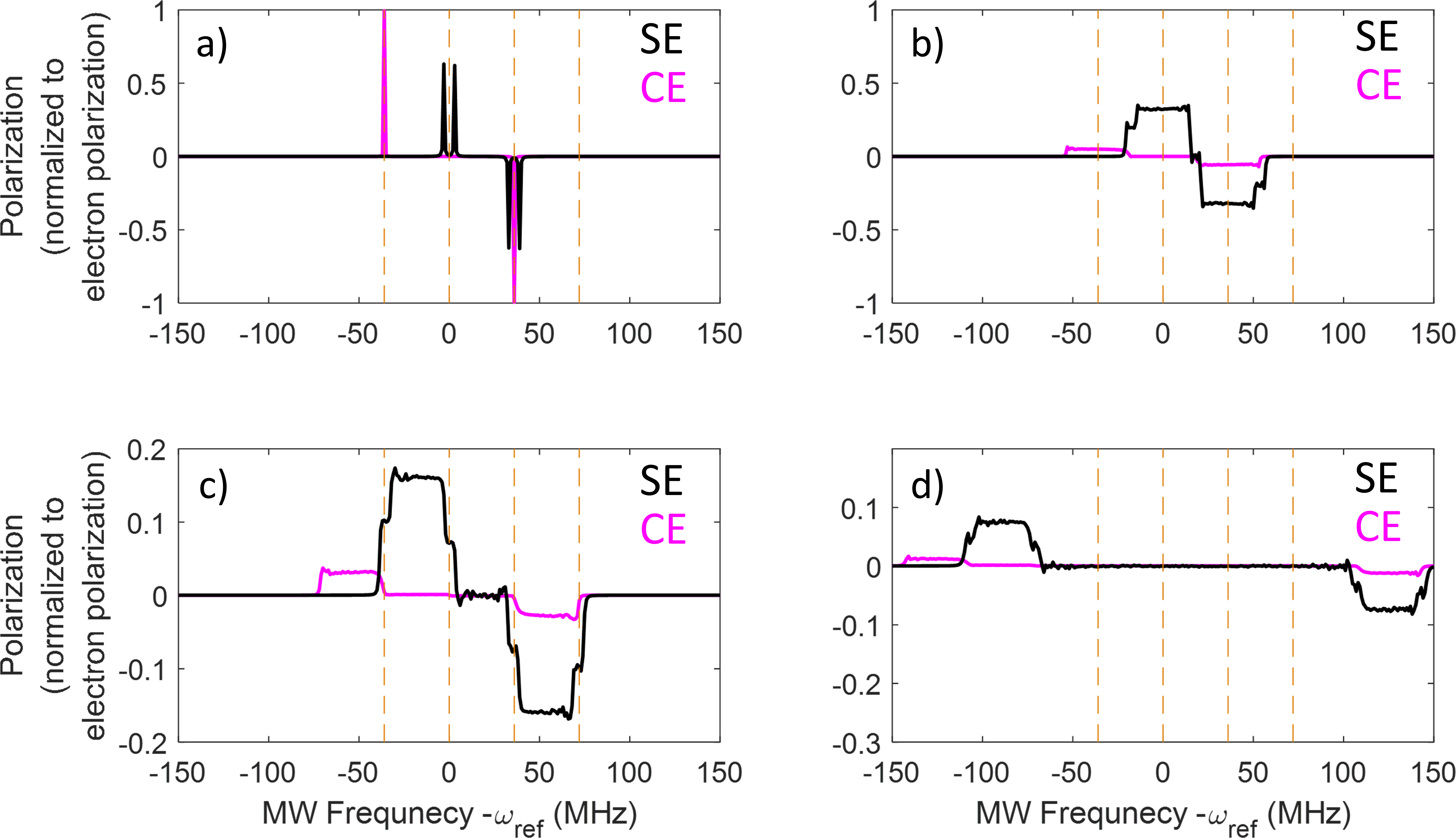}
    \caption{Simulated DNP spectra for an e--C system exhibiting the SE (red) and an e--e--C system exhibiting the CE (blue). Nuclear polarization (as a function of MW irradiation frequency, referenced to $\omega_{ref}=\omega_e = 94$ GHz. All polarizations are normalized to the electron polarization at thermal equilibrium. a) no modulation, b) $\Delta\omega=0.5\omega_C$, c) $\Delta\omega=\omega_C$, and d) $\Delta\omega=3\omega_C$. The dashed orange lines mark the position of the e--C DQ and ZQ transitions.
    The simulation parameters: \textbf{For SE (e--C):} $\omega_e=$94 GHz, $\omega_C=$36 MHz, $A_z^n=$0 MHz, $A_x^n=A_y^n=$0.5 MHz, $\omega_1=$0.5 MHz, T$_{1e}=$1.3 ms, T$_{1C}=$10 s, T$_{2e}=$10 $\mu$s, $T_{2C}=$100 $\mu$s, $t_\text{MW}=$100 s and $f_m=$50 kHz. No cross relaxation was added to the system.
    \textbf{For CE (e--e--C):} $\omega_{e1}=$94 GHz, $\omega_{e2}=$94.035876 GHz $\omega_C=$36 MHz, $D_{ee}=3$ MHz, $A_{z1}^C=$0 MHz, $A_{x1}^C=A_{y1}^C=$0.5 MHz, $A_{z2}^C=$0 MHz, $A_{x2}^C=A_{y2}^C=$0 MHz, $\omega_1=$0.05 MHz, T$_{1e1}=$T$_{1e2}=$1.3 ms, T$_{1C}=$10 s, T$_{2e1}=$T$_{2e2}=$10 $\mu$s, $T_{2C}=$100 $\mu$s, $t_\text{MW}=$100 s, $f_m=$50 kHz and T=$273$ K. No cross relaxation was added to the system.} 
    \label{fig:SE_CE_sims}
\end{figure}

The effect of frequency modulation on the SE- and CE-DNP enhancement in small spin systems was originally described by Hovav {\em et al}.\cite{Hovav2014}. Figure \ref{fig:SE_CE_sims} shows the effect of $\Delta\omega$ for the SE and the CE, at a fixed modulation frequency $f_m = 50$ kHz, for the diamond sample under study and the key features are summarized in Table \ref{tab:SE_CE_small}.

In our simulations, we simplify the diamond sample, and consider a single electron or a pair of electrons hyperfine coupled to a $^{13}$C nucleus. The addition of a $^{14}$N nucleus will just result in three EPR manifolds, each of which will behave independently, as we previously described \cite{diamond_1_2022_arxiv}. For an e--C system, the Hamiltonian of the spin system in the MW rotating frame is

\[H_{0}=\Delta\omega _{e}S_{z}-\omega _CI_z^C+A_{x}^CS_{z}I_x^C+A_{y}^CS_{z}I_y^C\]

\noindent where $S$ and $I^C$ are the electron and carbon spin operators, respectively. $\Delta\omega_e=\omega_e-\omega_{\text{MW}}$ is the electron off-resonance (compared to the rotating frequency $\omega_{\text{MW}}$), $\omega_C$ is the $^{13}$C Larmor frequency. $A_{x}^C$ and $A_{y}^C$ are the pseudo-secular terms of the dipolar hyperfine interaction. Note that no e-C secular hyperfine, $A_z^n=$0 MHz, was added to the simulation. Here, $A_{x}^C=A_{y}^C$.  For an e$_2$--e$_1$--C system the Hamiltonian is

\[H_{0}=\Delta\omega _{e1}S_{z1}+\Delta\omega _{e2}S_{z2}+D_{ee}S_{z1}S_{z2}-\omega _CI_z^C+A_{x}^CS_{z1}I_x^C+A_{y}^CS_{z1}I_y^C\]   

\noindent where $D_{ee}$ is the electron-electron dipolar interaction.

\begin{table}[htb]
\centering
\begin{tabularx}
{0.99\textwidth}{
      | >{\centering\arraybackslash}p{0.9 in}
      | >{\centering\arraybackslash}p{1.5 in}
      | >{\centering\arraybackslash}p{1.5 in}
      | >{\centering\arraybackslash}X |}
     \hline
      & Enhancement conditions & Enhancement when $\Delta\omega<\omega_C$ & Enhancement when $\Delta\omega\geq\omega_C$ 
      \\ \hline
     SE \newline e--C system 
     & 
     Positive or negative enhancement when irradiating at $\omega_e\pm\omega_C$, respectively \newline\newline (see Figure \ref{fig:SE_CE_sims}(a) red)
     & 
     unchanged \newline\newline (see Figure \ref{fig:SE_CE_sims}(b) red)
     & 
     sharp decline when $\omega_{e}-\Delta\omega<\omega_\text{MW}<\omega_{e}+\Delta\omega$  and a decrease of enhancement outside this range due to a reduction in the irradiation efficiency \newline\newline (see Figure \ref{fig:SE_CE_sims}(c-d) red)
     \\ \hline
     CE \newline 
     e$_{a}$-e$_{b}$-C \newline system \newline\newline
     $\omega_{e}^{a}<\omega_{e}^{b}$  $\omega_{e}^{b}-\omega_{e}^{a}\approx\omega_C$ 
     & 
     Electron polarization difference results in nuclear polarization enhancement according to $P_{C}=\frac{P_{b}-P_{a}}{1-P_{b}P_{a}}$ \newline\newline (see Figure \ref{fig:SE_CE_sims}(a) blue)
     & 
     slight increase due to the ability to irradiate on both sides of the EPR line that is split by the electron-electron dipolar coupling \newline\newline (see Figure \ref{fig:SE_CE_sims}(b) blue)
     & 
     sharp decline as a result of partial/full saturation of both CE-electrons and a drop in the CE enhancement even when only a single electron is irradiated due to a reduction in the irradiation efficiency \newline\newline (see Figure \ref{fig:SE_CE_sims}(c-d) blue)
     \\ \hline
    \end{tabularx}
    \caption{Description of the enhancement for the SE and the CE under different modulation amplitude conditions.  $\omega_e$ and $\omega_C$ are the electron and $^{13}$C Larmor frequencies.}
    \label{tab:SE_CE_small}
\end{table}

    In the original work by Hovav et al.\cite{Hovav2014}, they showed that it is best if the modulation rate (the inverse of the frequency of modulation, $1/f_{m}$) is set to be smaller than the electron relaxation time, T$_{1e}$ so that the electron spin does not lose its polarization between irradiation periods.\cite{Hovav2014} This is true for both the SE and the CE.

    The modulation amplitude, $\Delta\omega$, depends on the sample under study, and specifically on which nucleus we would like to enhance. For the SE, the MW irradiation $\omega_\text{MW}$ must be placed on the double quantum (DQ: $\omega_e-\omega_C$) or zero quantum (ZQ: $\omega_e+\omega_C$) transition in order to achieve positive or negative DNP enhancement, respectively (Figure \ref{fig:SE_CE_sims}a red).  At modulation amplitudes of $\Delta\omega<\omega_C$ the DNP enhancement is unchanged (Figure \ref{fig:SE_CE_sims}b red). However, once $\Delta\omega\geq\omega_C$, there is a sharp decline in the DNP enhancement of the SE in the range of $\omega_{e}-\Delta\omega<\omega_\text{MW}<\omega_{e}+\Delta\omega$ (Figure \ref{fig:SE_CE_sims}c-d red). This is a result of the irradiation saturating the electron single quantum (SQ) electron transition (at $\omega_{e}$) in addition to the e--C DQ or ZQ transition. When this occurs, there is a decrease in the electron polarization that can be transferred to the nucleus. In addition, the SE enhancement outside of the frequency range given above drops due to a reduction in the irradiation efficiency at each MW frequency, with the increase in the modulation amplitude $\Delta\omega$.
    
    For the CE, we must have two electrons at the CE condition (two electrons separated by the nuclear Larmor frequency $\omega_C$). The polarization that is transferred to the nucleus is equal to the polarization difference between the two electrons. When irradiating on the SQ transition of the low frequency electron, positive enhancement is achieved, and on the SQ transition of the high frequency electron negative enhancement is achieved (Figure \ref{fig:SE_CE_sims}a blue). At low modulation amplitudes of $\Delta\omega<\omega_C$ there is an increase in the CE enhancement due to the ability to irradiate on both sides of the EPR line that is split by the electron-electron dipolar coupling (Figure \ref{fig:SE_CE_sims}b blue). However, once $\Delta\omega\geq\omega_C$, there is a sharp decline in the DNP enhancement of the CE as a result of the irradiation saturating both CE-electrons, resulting in no polarization difference available to be transferred to the nucleus (Figure \ref{fig:SE_CE_sims}c-d blue). In addition, as $\Delta\omega$ is increased, there is a drop in the CE enhancement even when only a single electron is irradiated due to a reduction in the irradiation efficiency. The effective irradiation intensity depends on the strength of the state mixing, and is of the order of $2^{-1/2}\omega_{1}$ \cite{Hovav2014} where $\omega_1$ corresponds to the strength of applied MW field. Therefore, in order to observe the effect of decreasing MW irradiation efficiency under the CE conditions, Hovav et al.\ performed the simulations at low MW field strengths of 30 kHz \cite{Hovav2014}. The CE simulations shown in Figure \ref{fig:SE_CE_sims} were performed at $\omega_1/2\pi = 50$ kHz.

\subsubsection{Large spin system simulations}
\begin{figure}[htb]
    \centering
    \includegraphics[width=0.5\textwidth]{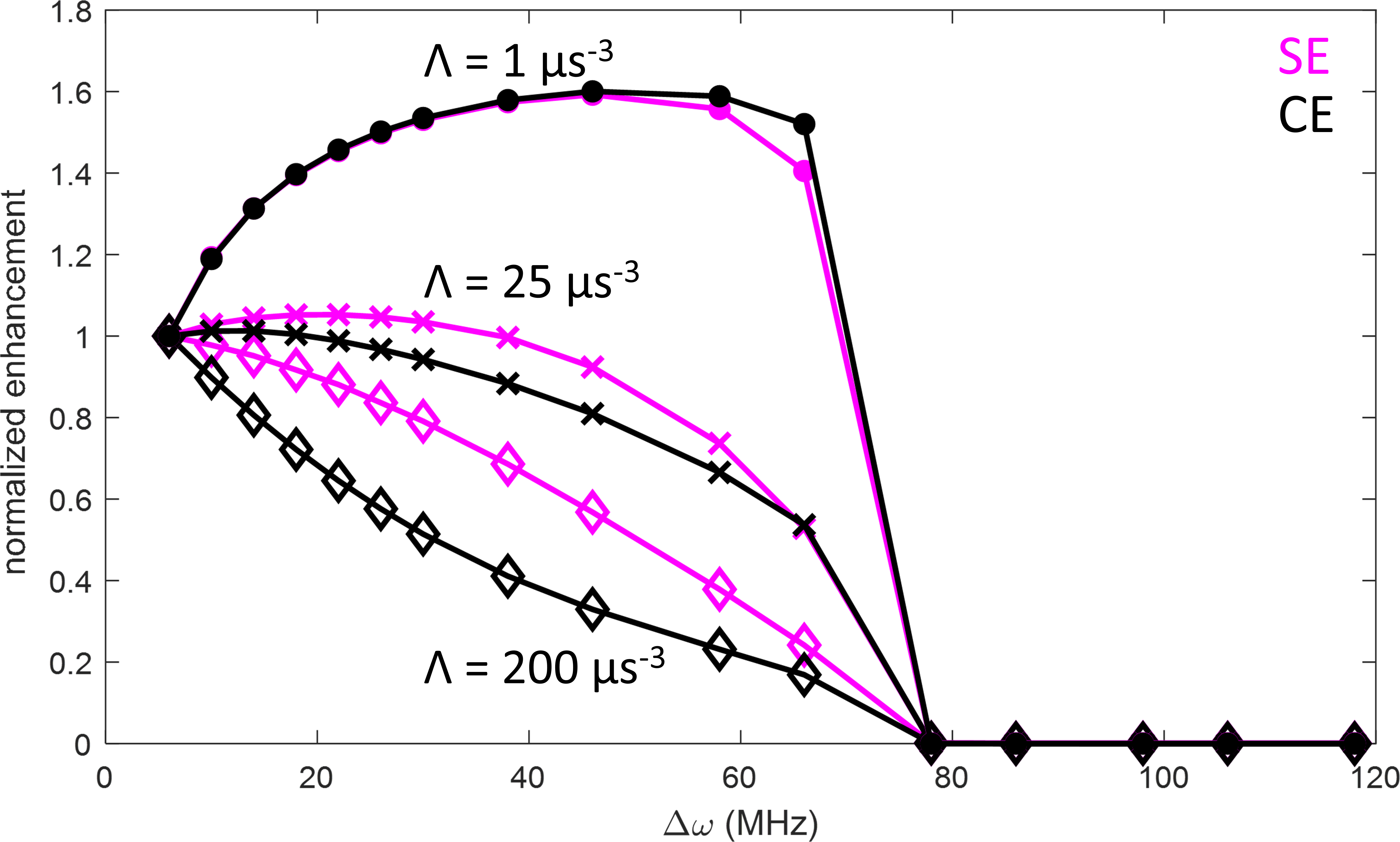}
    \caption{The SE (red) and CE (blue) enhancement of a single electron bin as a function of the $\Delta\omega$ for various electron spectral diffusion values ($\Lambda^{eSD}$ -shown in the figure). 
    The simulation parameters:
    The simulations were done using a model Gaussian EPR line with a full width at half max of 40 MHz, centered around     $\omega_e=94$ GHz. $\omega_C=36$ MHz, T=$273$ K, T$_\text{MW}=100s$, $\omega_1=0.5$ MHz, $\bar{A^{\pm}}=0.5 MHz$, T$_{1e}=1.3$ ms, T$_{1C}=10$ s and T$_{2e}=10 \mu$s.}
    \label{fig:large-spin-systems_SE_CE}
\end{figure}

The small spin simulation results do not align closely with the experimental data.
In order to describe a real sample, we must next consider how modulation affects the SE and the CE in large spin systems, when there are many electrons that are possible DNP sources. As noted earlier it is essential to consider the effects of eSD on the DNP process. In Figure \ref{fig:large-spin-systems_SE_CE} we consider the cases of (i) negligible eSD ($\Lambda = 1 \mu$s$^{-3}$ -- circles); (ii) intermediate eSD ($\Lambda = 25 \mu$s$^{-3}$ -- crosses); and (iii) large eSD ($\Lambda = 200 \mu$s$^{-3}$ -- diamonds).
In order to show the effect of the modulation amplitude without considering the addition of more electron spin packets as DNP sources, we plot the normalized DNP enhancement from a single electron spin packet in Figure \ref{fig:large-spin-systems_SE_CE}. This gives an indication to how a broad EPR line affects the DNP achieved from a single electron spin packet within the wider EPR line. 

\vspace*{0.15in}
\noindent {\em Negligible electron spectral diffusion}\\
For the SE (red circles), at low modulation amplitudes (below the optimum $\Delta\omega$), when irradiating at a given MW frequency $\omega_\text{MW}$, we are irradiating on the DQ and ZQ transitions of electrons that have a SQ transition at $\omega_\text{MW}\pm\omega_C$. However, we are also partially saturating the SQ transition of the electrons at $\omega_\text{MW}$. Partial saturation of these electrons results in a low DNP enhancement. As we increase the modulation amplitude $\Delta\omega$, the saturation of these SQ transitions becomes less effective, and we observe an increase in DNP enhancement. When we reach $\Delta\omega\approx\omega_{C}$, where we begin saturating both the DQ/ZQ transition and the SQ transition, and this causes a sharp decrease in the DNP enhancement. This results in an optimum $\Delta\omega\approx\omega_C$ (or slightly larger).
The CE (blue circles) behaves in a similar manner to the SE. Also here, the optimum modulation amplitude is $\Delta\omega\approx\omega_C$.

\vspace*{0.15in}
\noindent {\em Intermediate electron spectral diffusion}\\
When eSD is non-negligible, the electron depolarization spreads throughout the EPR line even when $\Delta\omega<\omega_C$, and saturates additional electron transitions that we are not directly irradiating.\cite{Shimon2022} As a result, we observe a lower optimum $\Delta\omega$ frequency, for the SE and the CE.\cite{Shimon2022} In the case of the CE, there is an increase because the eSD spreads the polarization throughout the line, so that even when we do not directly irradiate a CE electron we still observe its saturation.

\vspace*{0.15in}
\noindent {\em Large electron spectral diffusion}\\
At very large values of eSD, modulation does not improve the DNP contribution of a single electron to the SE or CE, and just results in a decrease in the enhancement. 

\vspace*{0.15in}
Going back to Figure \ref{fig:contours_SE_CE}, we are now in the position where we can discuss the effect of modulation on the SE and the CE. 
We see that SE is very sensitive to the combination of low modulation frequencies f$_m$ and large modulation amplitudes $\Delta\omega$. In this regime, the modulation rate ($1/f_{m}$) is likely on the order of, or larger than T$_{1e}$ making the effective MW power very low, and making it difficult to saturate the DNP transitions. Thus we observe a decrease in the enhancement. This is true in particular at large $\Delta\omega$ values, where the time spent irradiating each MW frequency is very short. It makes sense that the SE would be more sensitive than the CE, as the effective irradiation strength for the SE is already much weaker than for the CE.\cite{c3}

In our sample, we show that the SE enhancement only slightly improves with the application of modulation, and in most cases does not improve at all (Figure \ref{fig:contours_SE_CE}). This suggests that adding many electron spin packets to the DNP process is able to make up for the decrease in the SE enhancement from a single electron spin packet, such that the effects are balanced. Because we are observing the SE from the central EPR line, which is quite narrow (width at base of $\approx$ 20 MHz), increasing the modulation amplitude to values that are larger than the width of the central EPR line does not improve the DNP, but rather decreases it, especially at small f$_m$ values.

Moreover, the fact that we do not see a large increase in the SE enhancement in Figure \ref{fig:contours_SE_CE} also suggests that our sample has non-negligible eSD, and perhaps even quite strong eSD. Otherwise, we would expect to find optimal modulation parameters that would result in an increase in the SE enhancement. This was observed in the TEMPOL sample recently. \cite{Shimon2022}. 

In the CE case, we do find optimal modulation parameters that increase the enhancement. Therefore, we can conclude that the addition of more electron spin-packets into the DNP process is strong enough to overcome the fact that each spin packet is contributing less to the overall enhancement. This was also observed in the TEMPOL sample recently. \cite{Shimon2022}. Increasing the modulation amplitude, and eSD likely effect the CE more than the SE because the CE is resultant from a broader EPR line. When the line is broader, exciting the full EPR line is difficult at the low powers we are using. Increasing the modulation amplitude and eSD can help make up for this.
We must also remember that at large modulation amplitudes many of the DNP mechanisms overlap, making it difficult to draw hard conclusions about their behaviour under frequency modulation (see Figure \ref{fig:DNPsweepfits}).

Overall, the results from the SE and the CE suggest that by carefully selecting modulation parameters it is possible to change the relative ratios of the various DNP mechanisms, in systems where many mechanisms are active.

\subsection{Overhauser Effect and Truncated Cross Effect}

\subsubsection{Experimental Results} 
Next, we proceeded to explore the effect of frequency modulation at the OE and the tCE frequencies. Once again we plot the change in the enhancement compared to the non-modulated case as a function of the modulation amplitude and frequency.

\begin{figure}[H]
    \centering
    \includegraphics[width=0.99\textwidth]{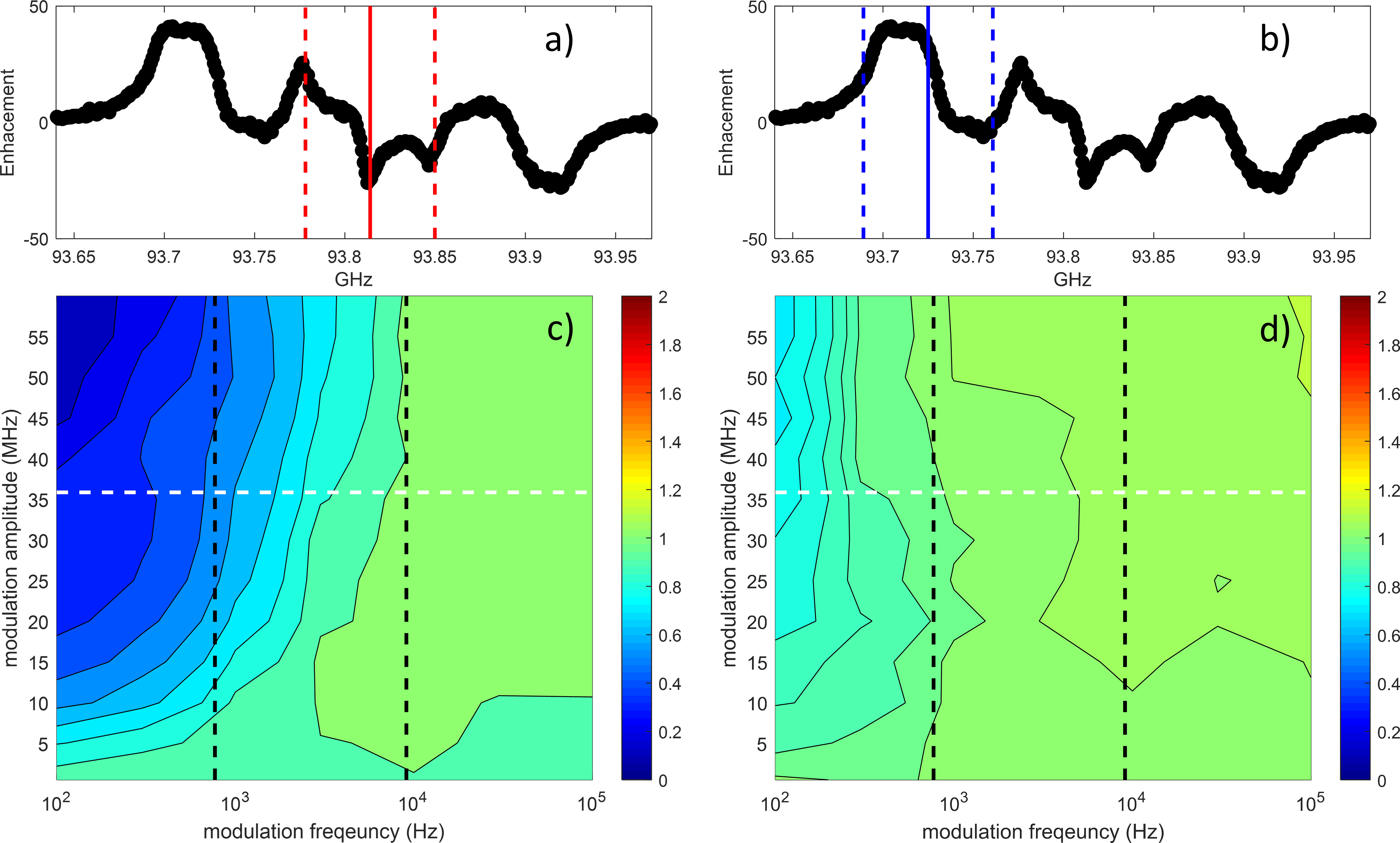}
    \caption{Non-modulated experimental DNP spectra showing the position of the central irradiation frequency(black symbols), $\omega_\text{MW}$, for the a) OE (solid red line) and b) tCE (solid blue line) mechanisms. The dashed blue lines represent the edges of the MW irradiation when $\Delta\omega=\omega_{C}$.
    DNP enhancement at the c) OE and d) tCE frequencies as a function of the modulation frequency, $f_{m}$, and the modulation amplitude, $\Delta\omega$. The dashed white line marks the position where $\Delta\omega=\omega_{C}$. The black dashed lines represent the two timescales of T$_{1e}$ values of a bi-exponential fit of an inversion-recovery experiment using pulsed EPR measured at 2.5 GHz, previously published.\cite{diamond_1_2022_arxiv} }
    \label{fig:contours_OE_tCE}
\end{figure}

Both mechanisms (but in particular the OE) are very sensitive to the combination of low modulation frequencies f$_m$ and large modulation amplitudes $\Delta\omega$ when enhancement is seen to drop sharply. Frequency modulation results in a slight increase for the OE (from -25.0 to -27.2) and a larger increase for the tCE (32.7 to 36.3) for large modulation frequencies and amplitudes.

\subsubsection{Small spin system simulations}
\begin{figure}[H]
    \centering
    \includegraphics[width=0.88\textwidth]{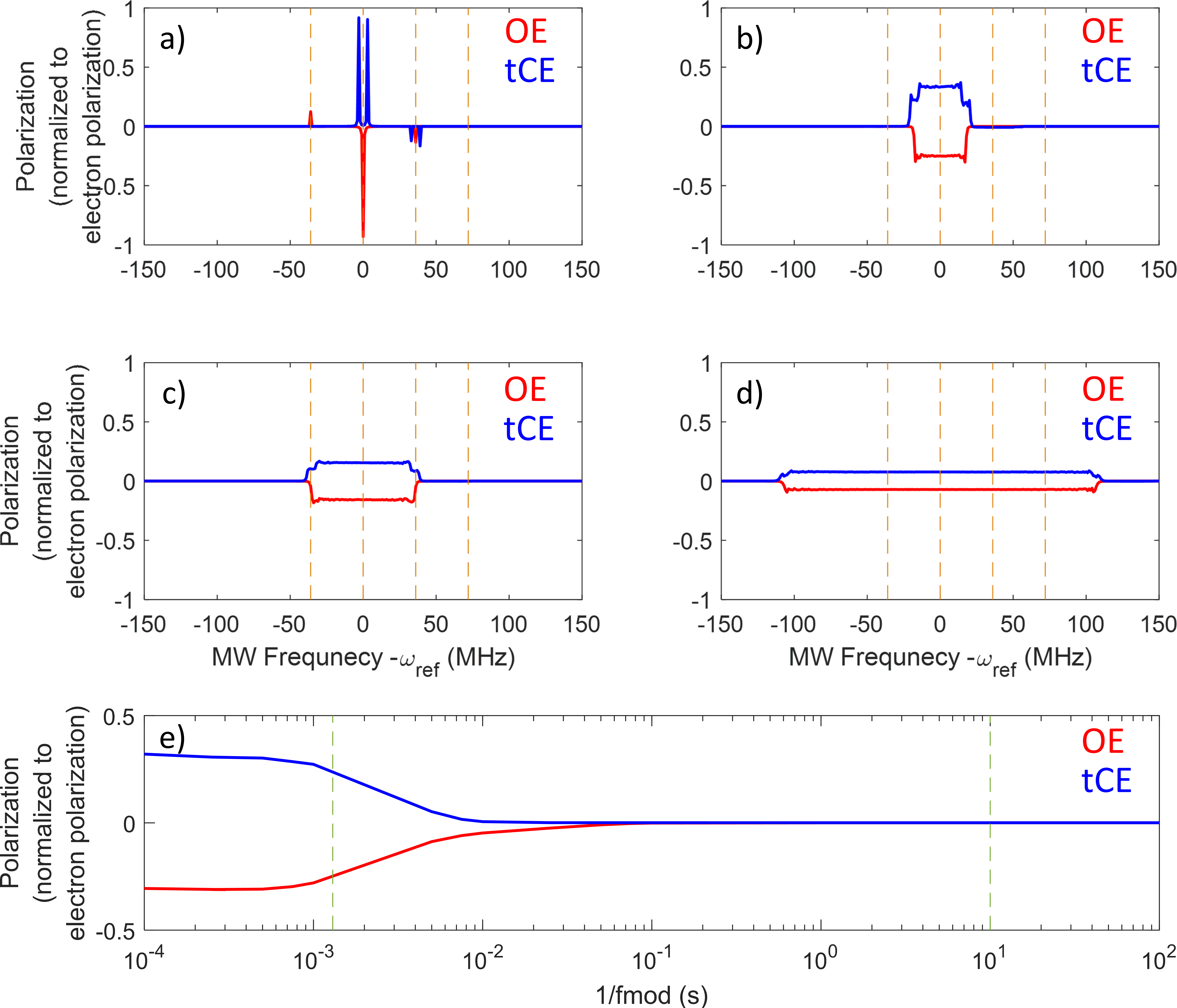}
    \caption{Simulated DNP spectra for an e--C system exhibiting the OE (black) and an e--e--C system exhibiting the tCE (magenta). Nuclear polarization (as a function of MW irradiation frequency, referenced to $\omega_{ref}=\omega_e = 94$ GHz. All polarizations are normalized to the electron polarization at thermal equilibrium. a) no modulation, b) $\Delta\omega=0.5\omega_C$, c) $\Delta\omega=\omega_C$, and d) $\Delta\omega=3\omega_C$. The dashed orange lines mark the position of the e--C DQ and ZQ transitions. e) Polarization as a function of $1/f_{m}$ for $\Delta\omega=\omega_C$. The green dashed lines represent 1/T$_{1e}$ and 1/T$_{1n}$.
    The simulation parameters: For OE (e--C): $\omega_e=$94 GHz, $\omega_C=$36 MHz, $A_z^n=$0 MHz, $A_x^C=A_y^C=$0.5 MHz, $\omega_1=$0.05 MHz, T$_{1e}=$1.3 ms, T$_{1C}=$10 s, T$_{2e}=$10 $\mu$s and $T_{2C}=$100 $\mu$s. $t_\text{MW}=$100 s, $f_m=$50 kHz and T=$273$ K. Cross-relaxation of T$_{1eZQ}=1000$T $_{1e}=$1.3 s and T$_{1eDQ}=10 $T$_{1e}=$13 ms were used.
    For tCE (e--e--C): $\omega_{e1}=$94 GHz, $\omega_{e2}=$94.035876 GHz $\omega_C=$36 MHz, $D_{ee}=3$ MHz, $A_{z1}^C=$0 MHz, $A_{x1}^C=A_{y1}^C=$0.5 MHz, $A_z2^C=$0 MHz, $A_x2^C=A_y2^C=$0 MHz, $\omega_1=$0.05 MHz, T$_{1e1}=$1.3 ms, T$_{1e2}= $0.013 ms, T$_{1C}=$10 s, T$_{2e1}=$T$_{2e2}=$10 $\mu$s and $T_{2C}=$100 $\mu$s. $t_\text{MW}=$100 s, $f_m=$50 kHz and T=$273$ K. No cross relaxation was added to the system.} 
    \label{fig:OE_tCE_sims}
\end{figure}

In order to better explain these observations, we expand the work of Hovav et al.\cite{Hovav2014} and describe the effect of frequency modulation on the OE and the tCE in small spin systems. The results are described in Figure \ref{fig:OE_tCE_sims} and summarized in Table \ref{tab:OE_tCE_small}.

\vspace*{0.15in}
\noindent {\em Overhauser Effect}\\
The simulated DNP spectrum with and without frequency modulation are shown in Figure \ref{fig:OE_tCE_sims}, for several $\Delta\omega$ values -- (a) no frequency modulation, (b) modulation with $\Delta\omega<\omega_C$, (c) $\Delta\omega=\omega_C$, and (d) $\Delta\omega>\omega_C$.
These spectra were calculated for an e--C spin system, and show a prominent negative OE-DNP mechanism when irradiating directly on the electron SQ transition. The e--C DQ and ZQ transitions are marked in the figure with dashed orange lines. At irradiation strengths of $\omega_{1}/(2\pi)=500$ kHz, the enhancement does not change as a function of the modulation amplitude (not shown), because of the large effective irradiation on the electron SQ transition. At lower irradiation strengths such as $\omega_{1}/(2\pi)=50$ kHz the enhancement decreases as the modulation amplitude increases (solid magenta lines), due to a reduction in the irradiation efficiency at each MW frequency, with the increase in the modulation amplitude $\Delta\omega$. In Figure \ref{fig:OE_tCE_sims}(a) a very small SE-DNP is seen when irradiating on the e--C DQ/ZQ transitions. It is small due to the weak effective irradiation, and is not visible in the other panels because the modulation further weakens it.

Next, we consider the effect of the frequency of the modulation on the DNP enhancement of the OE. As can be seen from Figure \ref{fig:OE_tCE_sims}(e), for the OE case, as in the SE and CE case, in order to effectively saturate the electron transitions it is best if $f_m$T$_{1e} > 1$ where T$_{1e}$ is marked by the dashed green lines. Here, as in the Hovav et al. paper, \cite{Hovav2014} we will not discuss the case where $1/(f_{m})<$T$_{2e}$.

\vspace*{0.15in}
\noindent {\em Truncated Cross Effect}\\
The simulated DNP spectrum with and without frequency modulation are shown in Figure \ref{fig:OE_tCE_sims}, for several $\Delta\omega$ values. These spectra were calculated for an e$_{1}$-e$_{2}$-n spin system, and show tCE-DNP when irradiating on e$_{1}$. The values of $\Delta\omega$ are given in the figure caption. They represent a) no frequency modulation, b) modulation with $\Delta\omega<\omega_C$, c) $\Delta\omega=\omega_C$, and d) $\Delta\omega>\omega_C$ conditions. The e--C DQ and ZQ transitions are marked in the figure with dashed orange lines. At irradiation strengths of $\omega_{1}/(2\pi)=500$ kHz, the enhancement does not change as a function of the modulation amplitude (not shown), because of the large effective irradiation on the electron SQ transition. At lower irradiation strengths such as $\omega_{1}/(2\pi)=50$ kHz, the enhancement decreases as the modulation amplitude increases (solid magenta lines), due to a reduction in the irradiation efficiency at each MW frequency, with the increase in the modulation amplitude $\Delta\omega$.
    
Finally, we consider the effect of the frequency of the modulation on the DNP enhancement of the tCE. As can be seen from Figure \ref{fig:OE_tCE_sims}e, for the tCE case, as in the SE and CE case, in order to effectively saturate the electron transitions we need $f_m$T$_{1e1} > 1$, where T$_{1e1}$ is the relaxation time of the of the slower relaxing electron. T$_{1e2}$ must be at least 100 times faster than T$_{1e1}$ for the tCE to be active (for slower T$_{1e2}$ values, we are in the range of CE). In addition, working at much larger modulation rates should be advantageous for the tCE. We will not discuss the case where $1/(f_{m})<$T$_{2e}$.

\begin{table}[htb]
\centering
\begin{tabularx}
{0.99\textwidth}{
 | >{\centering\arraybackslash}p{0.9 in}
      | >{\centering\arraybackslash}X
      | >{\centering\arraybackslash}X
      | >{\centering\arraybackslash}X |}
     \hline
     & Enhancement conditions & Enhancement when $\Delta\omega<\omega_C$ & Enhancement when $\Delta\omega\geq\omega_C$ 
     \\ \hline
     OE \newline
     e--C system 
     & 
     Positive or negative enhancement when irradiating at $\omega(e)$, depending on the relative ZQ/DQ cross relaxation rates \newline \newline (see Figure \ref{fig:OE_tCE_sims}(a) black)
     & 
     slight decrease due to a reduction in the irradiation efficiency \newline\newline (see Figure \ref{fig:OE_tCE_sims}(b) black)
     & 
     continued slight decrease due to a continued reduction in the irradiation efficiency \newline\newline (see Figure \ref{fig:OE_tCE_sims}(c-d) black)
     \\ \hline
     tCE \newline 
     e$_{a}$--e$_{b}$--C \newline system \newline\newline  $\omega_e^{a}<\omega_e^b$ and $\omega_e^{b}-\omega_e^{a}\approx\omega_C$ 
     & 
     Electron polarization difference results in nuclear polarization enhancement according to $P_{C}=\frac{P_{b}-P_{a}}{1-P_{b}P_{a}}$ \newline\newline (see Figure \ref{fig:OE_tCE_sims}(a) magenta)
     & 
     slight decrease due to a reduction in the irradiation efficiency \newline\newline (see Figure \ref{fig:OE_tCE_sims}(b) magenta)
     & 
     continued slight decrease due to a reduction in the irradiation efficiency \newline\newline (see Figure \ref{fig:OE_tCE_sims}(c-d) magenta)
     \\ \hline
    \end{tabularx}
    \caption{Description of the enhancement for the OE and the tCE under different modulation amplitude conditions}
    \label{tab:OE_tCE_small}
\end{table}

\subsubsection{Large spin system simulations}

\begin{figure}[htb]
    \centering
    \includegraphics[width=0.5\textwidth]{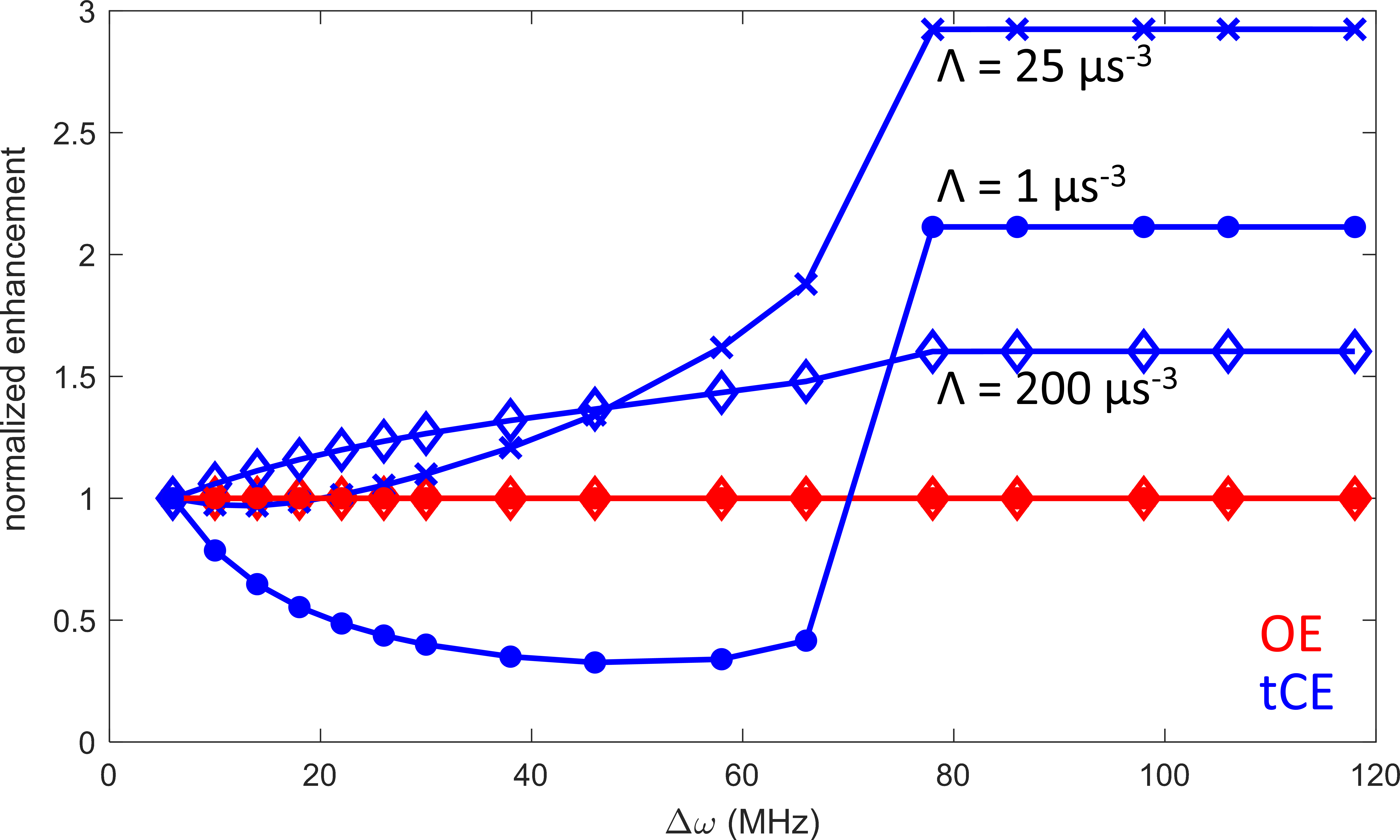}
    \caption{The OE (black) and tCE (magneta) enhancement of a single electron bin as a function of the $\Delta\omega$ for various electron spectral diffusion values (shown in the figure). 
    The simulation parameters:
    the simulations were done using a model Gaussian EPR line with a full width at half max of 40 MHz, centered around     $\omega_e=94$ GHz. $\omega_n=36$ MHz, T$=273$ K, T$_\text{MW}=100s$, $\omega_1=0.5$ MHz, $\bar{A^{\pm}}=0.5 $MHz, T$_{1e}=1.3$ ms, T$_{1n}=10$ s, T$_{2e}=10 \mu$s and $\Lambda^{eSD}$ as written in the figure}
    \label{fig:large-spin-systems_OE_tCE}
\end{figure}

In Figure \ref{fig:large-spin-systems_OE_tCE} we plot the effect of the modulation amplitude on the DNP enhancement from a single electron packet within a wider EPR line.  We again consider the cases of (i) negligible eSD ($\Lambda = 1 \mu$s$^{-3}$ -- circles); (ii) intermediate eSD ($\Lambda = 25 \mu$s$^{-3}$ -- crosses); and (iii) large eSD ($\Lambda = 200 \mu$s$^{-3}$ -- diamonds).

\vspace*{0.15in}
\noindent {\em Negligible electron spectral diffusion}\\
For the OE (black circles) the enhancement is not sensitive to the modulation amplitude. This is because the effective irradiation strength on the SQ electron transition is large even under large modulation amplitudes, $\Delta\omega$. Moreover, because we are irradiating on the SQ electron transition, the irradiation of the DQ/ZQ transitions does not affect the DNP enhancement, exactly as described above for small spin systems.
       
For the tCE (magenta circles), the enhancement is mediated via the polarization difference between the fast-relaxing pool of electrons (which is always fully polarized) and the slowly-relaxing pool of electrons. At low modulation amplitudes, the polarization difference is quite large because we are able to saturate the slow relaxing electrons. As the modulation amplitude increases, we see a decrease in the enhancement because the efficiency of the saturation of the slow electrons decreases as we spend less time irradiating at each frequency. However, at even larger modulation amplitudes we again see an increase in the enhancement. This comes from a second increase in the saturation of the slow electrons, as the modulation amplitudes become large enough to cover the whole width of the EPR line. When this happens, even irradiation that is not directly on the electron bin we are observing still results in saturation of that bin.
 
\vspace*{0.15in}
\noindent {\em Intermediate electron spectral diffusion}\\ 
When the eSD is non-negligible, the electron depolarization spreads throughout the EPR line, and saturates additional electron transitions that we are not directly irradiating even when $\Delta\omega<\omega_C$.\cite{Shimon2022} As a result, in the case of the tCE, there is an increase because the eSD spreads the polarization throughout the line, so that even when we do not directly irradiate on the slow relaxing electron pool, we still observe its saturation (see Figure \ref{fig:large-spin-systems_OE_tCE} -- magenta). The OE remains insensitive (all OE curves in Figure \ref{fig:large-spin-systems_OE_tCE} in black overlap)

\vspace*{0.15in}
\noindent {\em Large electron spectral diffusion}\\ 
At very large values of eSD we still observe a slight increase in the tCE enhancement, until the slow relaxing EPR line is fully saturated (see Figure \ref{fig:large-spin-systems_OE_tCE} -- magenta diamonds).

Going back to Figure \ref{fig:contours_OE_tCE}, we are now in the position where we can discuss the effect of modulation on the OE and the tCE. Looking at Figure \ref{fig:contours_OE_tCE}a-b more closely, we notice that the OE is very sensitive to the combination of low modulation frequencies f$_m$ and large modulation amplitudes $\Delta\omega$, similar to the SE. For the OE, the effective irradiation strength is much stronger because it relies on SQ irradiation. However, in this regime, the modulation rate ($1/f_{m}$) is likely of the order of, or larger than T$_{1e}$ making it difficult to saturate the DNP transitions, and thus resulting in a decrease in the enhancement. This is true in particular at large $\Delta\omega$ values, where the time spent irradiating each MW frequency is very short. 

For both the OE and the tCE, we get a slight increase in the enhancement (of the order of 10 \%) at large modulation amplitudes and large modulation frequencies. 
For the OE, the enhancement is a result of the narrow central EPR line, and therefore we would expect that increasing the modulation amplitude beyond the width of the central EPR line does not improve the DNP, but rather does not change it, or even decreases it, especially at small f$_m$ values. Note, also, that the DNP mechanisms overlap at large modulation amplitudes many of, making it difficult to draw hard conclusions about their behaviour under frequency modulation (see Figure \ref{fig:DNPsweepfits}).

For the tCE, the simulations of a single bin suggest that we should see an increase in the tCE enhancement. The simulations show that the increase in the tCE enhancement becomes smaller as the eSD becomes larger. This suggests that the slow electron pool in our system has a relatively large eSD connecting the electrons that make up the two outer EPR lines. This matches with our conclusions from the SE and CE enhancement described above. Thus, it seems like we are in the regime of large eSD, combined with a decrease in the effective irradiation power due to the modulation that decreases the tCE. We must also remember that the tCE overlaps with the CE and SE at large modulation amplitudes, making it difficult to separate the mechanisms.

\subsection{Conclusions}
Microwave frequency-modulation has recently been shown to improve the nuclear spin enhancements obtained during DNP.  While the effects of frequency-modulation on the SE and CE mechanisms have been studied previously, we have extended this analysis to the Overhauser effect (OE) and the truncated CE (tCE). We compare theoretical simulations of both small and large spin systems with experimental results from the the room-temperature $^{13}$C-DNP of diamond powders.  
We recently showed that DNP via P1 defects in diamond  can exhibit all of these mechanisms simultaneously due to the heterogeneity of P1 (substitutional nitrogen) environments within diamond crystallites \cite{diamond_1_2022_arxiv} - a property which we utilize here in this study.

We have shown that the overall resultant DNP enhancement observed in a sample during frequency modulation is a non-trivial combination of five factors, which we identified in this work:
(1) The larger the modulation amplitude, the more electron spins are involved in the enhancement process;
(2) The enhancement from a single electron bin as a function of modulation amplitude varies for different DNP mechanisms;
(3) The frequency of the modulation also affects the efficiency of the DNP;
(4) Which nucleus we are enhancing changes how the modulation affects each DNP mechanism; and
(5) The shape of the EPR line and especially its width compared to $\omega_n$ changes how the modulation affects each DNP mechanism.
By carefully picking modulation parameters, it is possible to suppress some mechanisms and enhance others, effectively changing how DNP proceeds in a very complex sample.

\section{Materials and Methods}

\subsection{Experiments}

\subsubsection{Sample preparation:}

Diamond powder used in these experiments were donated by Element6. The type Ib diamond is made by HPHT synthesis (High Pressure, High Temperature). The powder was 15-25 $\mu$m in size, and with a P1 concentration of 110-130 ppm.

\subsubsection{DNP spectrometer:}

The experiments were performed on a homebuilt DNP spectrometer, at a field of 3.4 T, corresponding to an electron Larmor frequency of 94 GHz, a $^{1}$H Larmor frequency of 142 MHz and a $^{13}$C Larmor frequency of 36 MHz. All experiments were performed at room temperature. The homebuilt DNP spectrometer and the MW bridge were described in detail recently.\cite{diamond_1_2022_arxiv}

The modulation amplitude reported in this work refers to half the amplitude of modulation, $\Delta\omega$, such that the full bandwidth of irradiation is: $\Delta\omega_{m} = \omega_{MW} \pm \Delta\omega$.
To achieve frequency modulation, a ramp-up sawtooth modulation was applied to the VCO, around a central value of 2.5 V. In order to convert from mV to MHz, a calibration was performed, using a spectrum analyzer to record the output of the voltage controlled oscillator (VCO) as a function of the voltage applied.

\subsubsection{$^{13}$C NMR and DNP experiments:}

The following settings were used for all the experiments, unless otherwise noted:

The DNP enhanced NMR signal was recorded using 20 or 24 scans, with a 90-acquire pulse sequence, using a $\pi$/2 pulse of 10 $\mu$s in length. Eight step phase cycling was used in all cases (no difference was found between 20 scans and 24 scans despite the phase cycle being incomplete). The thermal NMR signal was recorded using 100 scans. All experiments began with a train of 100 30 ms saturation pulses separated by 20 $\mu$s. We did not record the DNP enhancement at steady state. The length of the MW irradiation or the thermal recovery time after saturation was set to 120 s, as a compromise between the length of the experiment and the signal intensity.

Modulation grids were recorded by systematically changing the modulation amplitude and modulation frequency, while keeping all other experimental parameters unchanged.
The $^{13}$C relaxation time T$_{1n}$ and the enhancement buildup time T$_{bu}$ measurements were described in a previous publication.\cite{diamond_1_2022_arxiv}

\subsubsection{Data processing:}

Data processing was performed in MATLAB using custom scripts. 3 points of left-shift were used, in all cases. The data was then baseline corrected, phase corrected and 200 Hz exponential line broadening was applied. We report the integrated intensity of the $^{13}$C resonance. For DNP enhancement calculations we divide the integrated intensity of the MW-on signal with the MW-off signal. No enhancement is equal to 1.

\subsection{Simulations}

\subsubsection{Small spin systems}

Quantum mechanical simulations of a e--C and e$_{2}$--e$_{1}$--C spin systems which include relaxation effects and frequency modulation were performed according to the methods described by Hovav et al.\cite{Hovav2014} In order to simulate the OE, we added unequal cross-relaxation terms on the ZQ and DQ transitions, T$_{1ZQ}$ and T$_{1DQ}$, respectively. The addition of the cross relaxation to the relaxation superoperator was achieved by assuming fluctuations of S$^{+}$I$^{-}$ + S$^{-}$I$^{+}$ for the ZQ interaction and S$^{+}$I$^{+}$ + S$^{-}$I$^{-}$ for the DQ interaction, as described by Hovav et al.\cite{Hovav2014} S$^{\pm}$ and I$^{\pm}$ are the raising and lowering operators for the electron and the nucleus, respectively. In order to simulate the tCE, we repeated the CE simulations, but with unequal T$_{1e}$ values for the two electrons.

The simulations include: The electron and nuclear Larmor frequencies, $\Delta\omega_e$ and $\omega_C$, respectively. Secular and pseudo secular hyperfine interactions of A$_{z}$, and A$^{x}=$A$^{y}$, respectively. The electron and nuclear spin-lattice and spin-spin relaxation times, T$_{1e}$,  T$_{1C}$,  T$_{2e}$ and, T$_{2C}$, respectively. Electron-nuclear cross-relaxation on the DQ and ZE transition, T$_\text{1ZQ}$ and T$_\text{1DQ}$, respectively. MW irradiation at a frequency of $\omega_\text{MW}$, for a time $t_\text{MW}$, with a step size of $\Delta$t and of a strength $\omega_{1}$. Sawtooth modulation (ramp up) is employed, with a modulation amplitude of $\pm\Delta\omega$ surrounding the central irradiation frequency $\omega_\text{MW}$, a modulation frequency of $f_{m}$. No electron spectral diffusion is included in these simulations.

\subsubsection{Large spin systems}

In order to simulate a large spin system, we rely on a previously published model by Hovav et al. \cite{Hovav2014} which we updated to include modulation effects. \cite{Shimon2022} The model describes the changes in the polarization of the electrons in the sample using rate equations of their polarization. The polarization of a single hyperfine-coupled nucleus is also included, to mimic the average of all nuclei in the sample. The EPR line is split into bins of index j, with a frequency $\omega_j$, an amplitude $f_j$, and of width $\omega_\text{bin}$. The effect of frequency modulation with an amplitude $\Delta\omega$ surrounding a central frequency $\omega_\text{excite}$ is also taken into consideration. The model assumes modulation with a frequency that is larger then $1/T_{1e}$, but does not consider an explicit $f_m$ value. The polarization of each bin, $P_e(\omega_{j}; \omega_\text{excite}; \Delta\omega)$ is then described by coupled rate equations. We assume that the frequency modulation affects N bins, according to $N=(\Delta\omega)/\omega_\text{bin}$, and we scale the rates for the single quantum (SQ) and double/zero quantum (DQ/ZQ) irradiation according to the inverse of the amount of time the MW irradiation spends in each bin, $\omega_\text{bin}/\Delta\omega$. For every bin, we include SQ irradiation, DQ/ZQ irradiation, $1/T_{1e}$ relaxation, $1/T_{1n}$. We also include electron spectral diffusion (eSD) which spreads polarization between every pair of bins j and j'. The spectral diffusion constant is described by $\lambda^{eSD}$, with units of $\mu s^{-3}$, and is a phenomenological parameter. All the rates are included in a single rate matrix $R$.

The rate equations for the electron polarization across the EPR line (which has the length of the number of bins N) and the polarization of a single nucleus are solved by calculating:

\begin{gather}
    \begin{bmatrix}
    \overrightarrow{P_e(\Delta\omega;t_\text{MW})} \\
    P_n(\Delta\omega;t_\text{MW}) 
    \end{bmatrix}
    =R
    \begin{bmatrix}
    \overrightarrow{P_e(\Delta\omega; 0)} \\
    P_e(\Delta\omega; 0) 
    \end{bmatrix}
\end{gather}

\noindent{}where the electron polarization across the EPR line, $\overrightarrow{P_e(\Delta\omega;t_\text{MW})}$, is represented by a vector of length N, which matches the number of bins in the EPR line, N.
The electron polarization and the shape of the EPR line are then used to simulate the DNP spectra for the various DNP mechanisms. The nuclear polarization is not used further, but is necessary to reproduce the effect of the hyperfine interaction on the electron polarization during the MW irradiation. The details of the SE and CE under frequency modulation were published previously.\cite{Hovav2014} This model in conjunction with frequency modulation was used to analyze experimental data for model samples with nitroxide radicals \cite{Shimon2022}. In this work we expand the model to include the OE and the tCE.

\paragraph{The SE:}
    
    \begin{multline}
        S_{SE}(\omega_\text{MW})=
        \sum_{\omega_\text{MW}}\Bigg( \sum_{\Delta\omega_\text{mod}=-\Delta\omega/2}^{\Delta\omega/2}\big( g(\omega_\text{MW}-\omega_n)P_e(\omega_\text{MW}-\omega_n+\Delta\omega_\text{mod},\omega_\text{MW}) 
        \\
        -g(\omega_\text{MW}+\omega_n)P_e(\omega_\text{MW}+\omega_n+\Delta\omega_\text{mod},\omega_\text{MW})\big)\Bigg)
    \end{multline}
    
    Where $\omega_\text{MW}$ is the central MW irradiation frequency, $\Delta\omega_\text{mod}$ is position of the irradiation around the central irradiation frequency, and goes from $-\Delta\omega/2$ to $\Delta\omega/2$ in steps that match the bin size, $\omega_n$ is the nuclear Larmor frequency, $P_e(\omega_x,\omega_y)$ is the electron polarization at a position $\omega_x$, when irradiating at a frequency $\omega_y$.\cite{Hovav2014}

\paragraph{The CE:}
    
    \[
    \sum_{\omega_\text{MW}}\left(
    g(\omega_\text{MW})g(\omega_\text{MW}-\omega_n)
    \frac{\left( P_e(\omega_\text{MW}-\omega_n,\omega_\text{MW})-P_e(\omega_\text{MW},\omega_\text{MW}) \right)}{\left(  1-P_e(\omega_\text{MW}-\omega_n,\omega_\text{MW})P_e(\omega_\text{MW},\omega_\text{MW})\right)}
    \right)
    \]

\paragraph{The OE:}
    
    \[
    S_{OE}(\omega_\text{MW})=
    \sum_{\omega_\text{MW}}\left(
    \sum_{\Delta\omega_\text{mod}=-\Delta\omega/2}^{\Delta\omega/2}
    g(\omega_\text{MW}+\Delta\omega_\text{mod})(P_e^0-P_e(\omega_\text{MW}+\Delta\omega_\text{mod},\omega_\text{MW})) 
    \right)
    \]
    
    Where $P_e^0$ is the electron polarization at thermal equilibrium (i.e., fully polarized).
    Here, we assume that the polarization that is transferred to the nucleus matches the amount of polarization that the electron loses during the DNP, in order to simulate negative enhancement. 
    
\paragraph{The tCE:}
    
    \[
    \sum_{\omega_\text{MW}}\left(
    g(\omega_\text{MW})g(\omega_\text{MW}-\omega_n)
    \frac{\left( P_e(\omega_\text{MW}-\omega_n,\omega_\text{MW})-P_e^0 \right)}{\left(1-P_e(\omega_\text{MW}-\omega_n,\omega_\text{MW})P_e^0\right)}
    \right)
    \]
    Here, we assume that there are two electron pools, one at $\omega_\text{MW}-\omega_n$ that is partially saturated, and another at $\omega_\text{MW}$ that is fully polarized. In order to achieve that, we always keep the polarization of the second electron equal to the thermal electron polarization $P_e^0$. As a result we assume that there are two identical EPR lines, one for a slow relaxing electron pool, and one for a fast relaxing electron pool that cannot be saturated.

\subsection{EasySpin Simulation of the EPR line}

The EPR line was simulated via EasySpin, as described in our previous publication. \cite{diamond_1_2022_arxiv}

\subsection{Fitting the Modulation DNP Spectra}

The DNP spectra with modulation were fitted using a crude method of convoluting the EPR line with delta functions to form the basic shapes for the SE, OE (same shape used for tCE) and CE DNP mechanisms, and changing their amplitude to achieve the best agreement with the experimental spectrum. This method was previously described in Banerjee et al. \cite{Banerjee2013}, for the SE and the CE, and for the OE and tCE in our previous publication on diamond DNP.\cite{diamond_1_2022_arxiv}

We added the effect of ramp-up (sawtooth) frequency modulation by convoluting the shapes for the SE, CE, OE and tCE with square functions. The width of these square functions matches the width of $\Delta\omega$ used in the experiment.

\section{Acknowledgements}

The authors thank Element 6 for the donation of the powder samples used in these experiments. This work was partially supported by funding from the National Science Foundation under cooperative agreement under grants OIA-1921199.

\bibliography{references}

\end{document}